# ENERGY EFFECTIVE MASS DEPENDENCE OF ELECTRON TUNNELING THROUGH CDS/CDSE, AL$_x$GA$_{1-x}$AS/GAAS AND ALSB/INAS MULTIPLE QUANTUM BARRIERS

## Jatindranath Gain[1], Madhumita Das Sarkar[2] and Sudakshina Kundu[3]

[1]Department of Physics, Derozio Memorial College, India
[2,3]Department of Electronics and Communication Engineering, Maulana Abul Kalam Azad University of Technology, India

*Abstract*

*Tunneling of electrons through the barriers in heterostructures devices is investigated by using the unified Transfer Matrix Method. The effect of barrier width on electron transmission coefficients has also been examined for different pairs of semiconductor devices of significant research interest in current years. Such Pairs involve: AlxGa1-xAs/GaAs, AlSb/InAs, and CdS/CdSe quantum barriers with varying dimensions reduced from 20 nm to 5nm to observe how tunneling properties are affected by scaling. The effective electron masses in the well and barrier regions typically vary with constituent materials. It has been shown that the transmission coefficients are significantly changed due to the coupling. The effective mass dependent transmission coefficients for electron energy have been evaluated in terms of the mass discontinuity metrics. The electron transmission coefficients for each pair of quantum structures are plotted with the variation of its electron energy, normalized to its potential energy. The resonant state obtained here will be beneficial for designing detectors, optical filters, photonic-switching devices and other optoelectronic and photonic devices.*

## 1. INTRODUCTION

In recent years, low-dimensional semiconducting transmission systems have acquired great relevance because their unique properties make them useful in applications ranging from optoeltronics to high-speed device [1], [2]. The carriers' perpendicular transit in semiconductor hetero-structure has garnered a lot of interest in this connection. Especially because of its potential utilization in the design as well as fabrication of single electron tunneling transistor, resonant tunneling diodes, resonant photo detectors, quantum cascade lasers etc., the MQW structures become extremely significant [3]-[5]. The carriers' tunneling effect becomes extremely significant with the reduction in the CMOS devices dimensions for estimation of the different leakage current which flows in the VLSI chips device [6], [7]. In this connection Quantum mechanical tunneling through multiple quantum barriers is a potential research interest [8].

In this paper, an attempt has been made to study carriers tunneling through a multiple quantum barrier and the transport coefficient variations have been plotted as a function of the carrier energy. The energy range also includes the transitions which are forbidden classically.

The established techniques for computation of energy splitting and tunneling probability are (i) the Numerical Calculation (ii) the Wentzel–Kramer's–Brillouin (WKB) Approximation and (iii) Instanton Method. The Instanton approach provides physical insights into quantum tunneling, but its validity is limited in case of large separation of the two potential minima. [9]. WKB approximation is frequently utilized in its basic mathematical form, but due to an inherent flaw in the connection formula the outcome is not accurate [10], [11]. WKB approximation has recently been produced by altering the loss phase at the traditional turning places, although it has given the ideal result in certain cases [12]. The Numerical approaches may yield results up to the necessary precision but at the cost of loss of a substantial amount of physical knowledge. [13]. This paper develops a model for the development of multiple quantum barriers or wells potential which is based on analytical Transfer Matrix Method (ATMM) that can be employed to any general and arbitrary asymmetric aperiodic MQW structures effectively. The model is used for three different devices and satisfactory results are obtained.

The coefficient of tunneling/transmission, which is the ratio of the flux of particles/carriers through the potential barriers to particle incidents flux at the other interfaces. It has been calculated using of Ben Daniel-Duke (BDD) boundary conditions in order to resolve the Schrodinger electron equations (in this case, the carriers) within coupled well areas as well as the barrier between them [14]. The hypothesis depends on the TMM method. Tunneling relies considerably on the width of the barrier. The structure dimension scaling greatly impacts this variation. The material pairs considered are InAs/AlSb/InAs, GaAs/Al$_x$Ga$_{1-x}$As/GaAs and CdSe/CdS/CdSe multiple quantum barriers.

HEMTs based on AlSb/InAs are suitable for satellite applications because of their low operating voltage [15-16]. Al$_x$Ga$_{1-x}$As /GaAs-based devices have been used for quite a while. But for light emitting applications CdS/CdSe QW architectures offer better profitability [17]. The effective mass can fluctuate with energy as in the AlSb/InAs pair. This will impact the tunneling behavior of the electrons as the transmission coefficient is dependent on effective mass. So, the effective mass variation of the quantum transport coefficient is also studied in this paper. The transport coefficient variation with electron power is investigated as well as for each of these material pairs for different barrier widths.

## 2. THEORY

The quantum mechanical tunneling theory via a traditionally prohibited energy state may be extended to various sorts of conventionally prohibited transitions [18]. In this study a quantum tunneling widespread theory has been created and tested for three distinct material pairs for transition to multiple quantum barriers. These neighboring lower energy areas, separated by quantum barrier, are linked together. This is the overall pattern for several heterostructures. In the lower energy regions and barrier area, electron wave functions are obtained by solving the Schrodinger equations with appropriate boundary requirements [14]. In the regions considered, the solutions depend on the carrier effective masses. Therefore, the tunneling probability will not only depend on the barrier size alone, but also on the changes of materials and their energy-dependent effective mass.

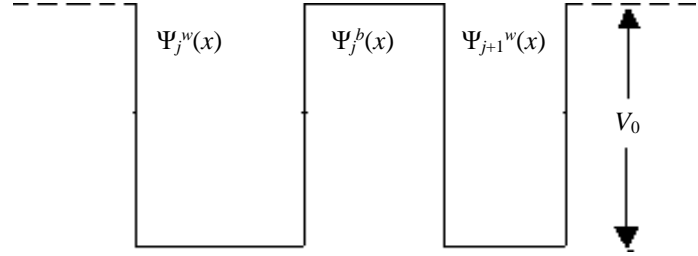

Fig.1. Schematic Representation of a "MQW (multiple quantum wells)" heterostructures

For the electron, total wave function $\Phi(r,x)$ is given by:

$$\Phi(r,x) = \exp(ik_0 r)\Psi(x) \qquad (1)$$

For the one-dimensional (1-D) Schrödinger equation, with $\Psi(x)$ for position-dependent electron effective mass and growth direction of heterostructure's represented by $x$. As per the effective mass theory, the Schrodinger Eq.(2) for the well regions on either side and the finite potential barrier takes the recognized form:

$$\frac{\hbar^2}{2m^*}\frac{d^2\Psi(x)}{dx^2} + (E-V)\Psi(x) = 0 \qquad (2)$$

With potential energies $V$ as well as suitable effective masses $m^*$ for the region wherein the Eq.is described. The potential energy is $V_0$ and the effective mass is $m^* = m_B$ inside the barrier area whereas it is represented by zero and $m^* = m_w$ respectively outside the barrier. Also, for the $j^{th}$ junction transfer matrix is utilized as well as for N-junctions, the result is generalized. Wave function coefficients are related through this matrix at the junction's one end to other end, thus the wave function represented as

$$\Psi_j^w(x) = A_j^w \exp(ik_w x) + B_j^w \exp(-ik_w x) \;\forall j^{th} \text{ well and} \qquad (3)$$

$$\Psi_j^b(x) = A_j^b \exp(ik_b x) + B_j^b \exp(-ik_b x) \;\forall j^{th} \text{ barrier.} \qquad (4)$$

where $k_w = \left(\frac{2m_w E}{\hbar^2}\right)^{0.5}$ and $k_b = \left(\frac{2m_w(V_0-E)}{\hbar^2}\right)^{0.5}$

With electron effective masses represented by $m_w$ as well as $m_B$ which are the in the outside barrier regions as well as barrier region respectively. Matching the wave function $\Psi(x)$ as well as its suitable normalised derivative:

Continuity $\frac{1}{m^*}\frac{d^2\Psi(x)}{dx^2}$ at the boundary as well as for each interface and for each interface 2×2 transfer matrix equations are obtained along with a matrix formula which relate the coefficient $A_j$ as well as $B_j$ with $A_{j+1}$ and $B_{j+1}$.

$$\begin{bmatrix} A_j \\ B_j \end{bmatrix} = \begin{bmatrix} M_{11}^{[j]} & M_{12}^{[j]} \\ M_{21}^{[j]} & M_{22}^{[j]} \end{bmatrix} \begin{bmatrix} A_{j+1} \\ B_{j+1} \end{bmatrix} \qquad (5)$$

The wave function's coefficients are obtained by utilizing Eq.(5) at the leftmost slab instead of the right most slabs.

$$\begin{bmatrix} A^j \\ B^j \end{bmatrix} = 0.5 \begin{bmatrix} 1 & -ik_w^{-1} \\ 1 & ik_w^{-1} \end{bmatrix} M_j \begin{bmatrix} 1 & 1 \\ ik_w^{-1} & -ik_w^{-1} \end{bmatrix} \begin{bmatrix} A_{j+1} \\ B_{j+1} \end{bmatrix} \qquad (6)$$

where $M_j$ represents the $j^{th}$ junction corresponding $j^{th}$ transfer matrix that is written as:

$$M_j = M_b(b_j)M_w(a_j)M_b(b_{j+1}) \qquad (7)$$

where $a_j$ and $b_j$ are the $j^{th}$ well and $j^{th}$ barrier's widths respectively, $M_w(a_j)$ and $M_b(b_j)$ represents the transfer matrices for the $j^{th}$ well as well as barrier respectively.

Herein, the total transfer matrix is stated as individual well and barrier's series cascading. The transmission amplitude $Q$ has been clearly discovered from Eq.(6) and Eq.(7) and represented as:

$$Q = \frac{1}{M_{11}+M_{22}+i(k_w M_{12}-k_w^{-1}M_{21})} \qquad (8)$$

where $M_{ij}$ represents the total transfer matrix elements. From the transmission coefficient ($T$) definition, we get

$$T = |Q|^2 \qquad (9)$$

The wave function's coefficients are obtained by utilizing Eq.(5) at the leftmost slab instead of the right most slabs

$$\begin{bmatrix} A_1 \\ 0 \end{bmatrix} = M^{[1]}M^{[2]}M^{[3]},...,M^{[j-1]} \begin{bmatrix} A_j \\ 0 \end{bmatrix} \qquad (10)$$

$$\begin{bmatrix} A_1 \\ 0 \end{bmatrix} = M_{total} \begin{bmatrix} A_j \\ 0 \end{bmatrix} \qquad (11)$$

which meets $M_{21}=0 \qquad (12)$

We achieved the energy eigen functions as well as eigen values through elucidating Eq.(12).

For every positive electron energy $E$, the wave vector $k_w$ outside the barrier is real. The lowest conductive band is the zero energy in the area outside the barrier. As the barrier's band of drive is above the area outside, the wave vector $k_b$ will be imaginary for the electron ($E<V_0$) certain energies as well as for energies above $V_0$ ($E>V_0$), real.

The energy is therefore split into 2 areas, $E>V_0$, where transition can also be carried out under classical circumstances and ($E<V_0$) for non-classical tunneling transition. The answers will vary and the methodologies used to estimate the transmission coefficients in both locations will vary. Since $k_w$ and $k_b$ both are effective mass and energy dependent, for energy values the variance in transmission coefficient, normalized in relation to the barrier height, will thus differ for various material combinations.

Herein, $\beta = m_B/m_w$ known as mass discontinuity factor which play a significant part in transmission coefficient determination. For a complete transmission via a sandwiched barrier among two wells, it is required that,

$$\left[\beta\left(\frac{k_b}{k_w}\right) - \frac{k_w}{\beta k_b}\right]^2 \sinh^2 k_w L = 0 \qquad (13)$$

where, $L$ represent the barrier length. Thus, at energy values where there will be transmission are:

$$E_n = V_0 + \frac{(n\pi\hbar/L)}{2m_B} \qquad (14)$$

where the transmission coefficient is unity. At these energy values, the barrier becomes transparent. In atomic physics, it is known as Ramsauer-Townse effect [19]. At normalised energy values, the resonance conditions are fulfilled.

$$\left(\frac{E}{V_0}\right) = 1 + \frac{n^2}{\beta} \quad (15)$$

where $n$ represents an integer. Thus, resonance values are mass discontinuity factor dependent.

When the barrier height is more than the electron energy $E$, this condition is rather different. It has been assumed by above derivation that with energy there is no change in the mass discontinuity factor. Though, this is highlighted that the variation in the effective mass with the energy for InAs/AlSb material system is as per the expression:

$$m(E) = m^*[1 + (E-E_c)/E_{eff}] \quad (16)$$

where, $E_c$ represents the conduction band minimum as well as $E_{eff}$ represents the effective band gap. The transmission coefficient usually depends on energy via the wave vector $k_b$ equation. This fluctuation is further compounded by the mass discontinuity factor $\beta(E)$ energy variation. The effect of this variance and how it impacts the tunneling phenomena are worth examining.

The electrons' transmission coefficient across a potential barrier is crucial for the investigation of the leakage current in Nanometer-based MOSFETs [20]. This is an important parameter for examining the behavior of many quantum well architectures, which sandwich the barrier among two coupling quantum wells. Another instance involves modifying the Eq.to reflect the well dimensions. If both barrier and well areas exist in the nanoscale range, it has been expected that the energy level to be further quantified. It is examined in the multiple quantum well structure's further examination.

## 3. RESULTS AND DISCUSSIONS

The numerical results acquired by utilizing MATLAB programming across multi-barrier heterostructure are presented. The selected material pairs are AlSb/InAs, Al$_x$Ga$_{1-x}$As/GaAs and CdS/CdSe. The Table.1 represents the parameters that were utilized in the computations. In Al$_x$Ga$_{1-x}$As the electron's effective mass depend on $x$'s mole fraction, wherein Al concentration represented by the $x$ [21] [22]. m$_{AlxGa1-xAs}$=(0.063+0.083$x$) represents the effective mass in Al$_x$Ga$_{1-x}$As case as well as E$_{AlxGa1-xAs}$= (1.9+0.125$x$+0.143$x^2$) represents the energy band gap $E$.

For every pair 5nm, 10nm and 20nm barrier width $L$'s three values are considered. In every example, the coefficient of transmission increases as predicted with a decreasing dimension. For InAs/AlSb, it is the slowest rise with the $\beta$'s highest value. The effective mass variation in AlSb/InAs pair is considered in the computation.

It is seen that for AlSb/InAs the mass discontinuity factor does not change much. Though, such slow variation of mass discontinuity factor $\beta$ seems to slow down the coefficient of transmission increase, particularly with the extremely low barrier width (5nm). In AlSb/InAs, quantum tunneling is found to be least for all barrier widths while it is high for the Al$_x$Ga$_{1-x}$As/GaAs pair, Al$_x$Ga$_{1-x}$As being most significant. It is justified very easily as GaAs/Al$_x$Ga$_{1-x}$As/GaAs structure has the lowest barrier height and InAs/AlSb/InAs having highest as shown in Fig.2-Fig.4.

There is an increase in transmission coefficient with decreasing barrier width as well as with normalised electron energy. For GaAs/Al$_x$Ga$_{1-x}$As/GaAs, there is sharp increase while for InAs/AlSb/InAs it is rather slower. It is clearly described in case of comparison of barrier heights. Also, CdSe/CdS/CdSe structures are found in halfway.

There is an increase in transmission coefficient in a non-linear fashion from 0 to 1 for ($E_{nor}=E/V_0$)<1. For every pair, the transmission coefficient for wider wells is low as predicted, although there is slow increase for InAs/AlSb/InAs due to the mass discontinuity factor's highest value.

There is resonance beyond standard energy ($E_{nor}=E/V_0$)>1, i.e., there are quantum energy values, when transmission achieves acute maximum levels. For all three structures, in particular around $E_{nor}$=1, this variation is noticeable. With decreasing barrier width, the peaks are separated more energetically; in the GaAs/AlGaAs/GaAs structure this is most obvious. For InAs/AlSb/InAs, the gap between maximum and minimum values remains largely stable for decreasing breadth. The maximum transmission coefficient progressively improves as the $E_{nor}$ increases after 1 for GaAs/AlGaAs/GaAs and CdSe/CdS/CdSe. Also, there is a gradual increase in the difference between the minima and maxima.

Herein, region of lower band-gap material is considered as semi-infinite. Therefore, there is continuous variation in the normalized energy. In case outside the barrier, lower band gap material's width decreases in nanometers order, it inside the quantum wells resulted in the quantization of energy values. Such effects with normalised energy are reflected in the transmission coefficients as well as considered to have a significance change. In field effect devices, leakage current and multiple quantum wells and barriers structures' carrier tunneling is crucially affected through it. The author explored the researches of the particular effect.

Table.1. Computational Parameters

| Parameters | CdS/CdSe | Al$_x$Ga$_{1-x}$As/GaAs ($x$=0.47) | AlSb/InAs |
|---|---|---|---|
| Conduction Band gap ($\Delta E_c$) | $Eg_{CdS}$=2.36eV<br>$Eg_{CdSe}$=1.69eV<br>$\Delta E_c$=67%×($Eg_{CdS}$- $Eg_{CdSe}$)<br>$\Delta E_c$=0.54eV | $Eg_{AlxGa1-xAs}$=2.36eV<br>$Eg_{GaAs}$=1.69eV<br>$\Delta E_c$=67%×($Eg_{AlxGa1-xAs}$- $Eg_{GaAs}$)<br>$\Delta E_c$=0.54eV | $Eg_{CdS}$=2.36eV<br>$Eg_{CdSe}$=1.69eV<br>$\Delta E_c$=67% of ($Eg_{CdS}$- $Eg_{CdSe}$)<br>$\Delta E_c$=0.54eV |
| Effective Mass ($m^*$) | $m^*_{CdS}$=0.20$m_0$<br>$m^*_{CdSe}$=0.13$m_0$ | $m^*_{AlxGa1-xAs}$=0.106$m_0$<br>$m^*_{GaAs}$=0.13$m_0$ | $m^*_{AlSb}$=0.098$m_0$<br>$m^*_{InAs}$=0.020$m_0$ |
| Mass Discontinuity $\beta=m_B/m_w$ | $\beta$=1.54 | $\beta$=1.58 | $\beta$=4.9 |

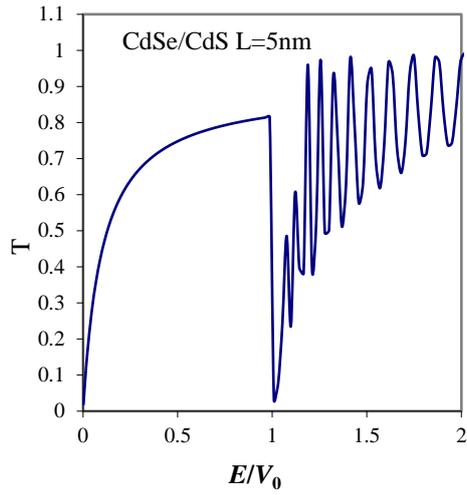

(a)

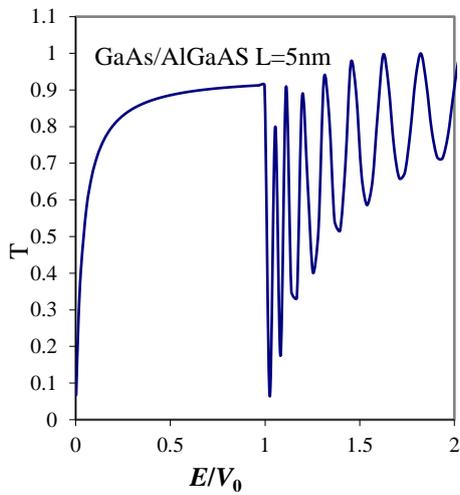

(b)

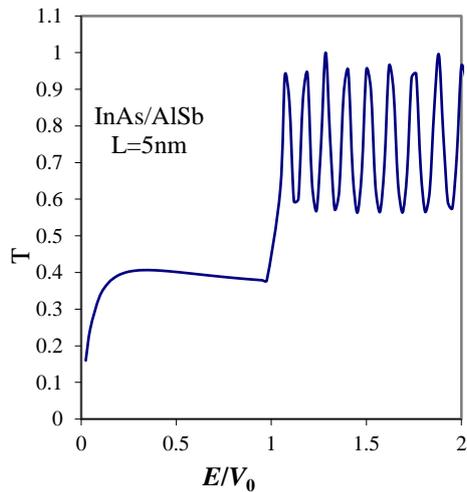

(c)

Fig.2. Variation of electrons' transmission coefficient ($T$) with normalized energy ($E/V_0$) for (a) CdS/CdSe (b) AlGaAs/GaAs (c) AlSb/InAs multiple quantum barriers (MQB) for barriers width ($L$) 5 nm

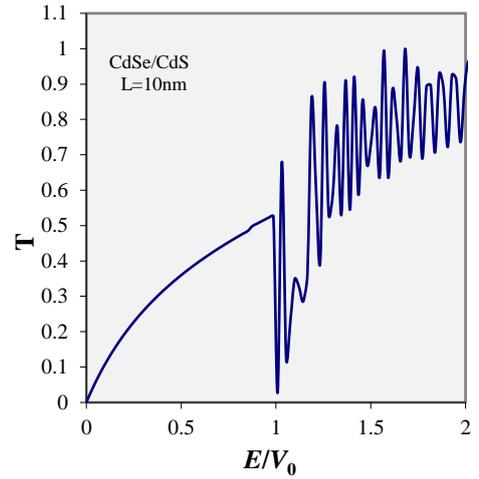

(a)

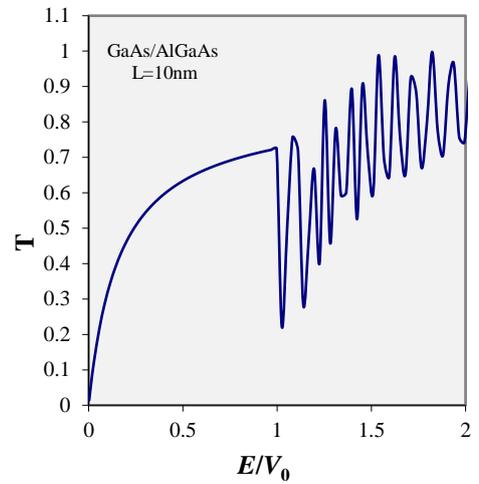

(b)

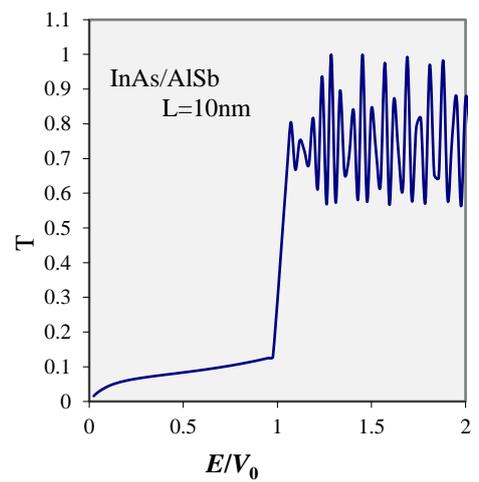

(c)

Fig.3. Variation of electrons' transmission coefficient ($T$) with normalized energy ($E/V_0$) for (a) CdS/CdSe (b) AlGaAs/GaAs (c) AlSb/InAs multiple quantum barriers (MQB) for barriers width ($L$) 10 nm

## 4. CONCLUSION

The established general model employed for any aperiodic MQW structures' designing. For achieving tunnelling at any specific energy value through the structure at the specified energy the widths of well for $\tau=1$ can be iteratively calculated. During the resonance tunnelling, the electron energy resonates at the bound states of the single quantum well. The Energy band profiles in CdS/CdSe, $Al_xGa_{1-x}As$/GaAs and AlSb/InAs Multiple Quantum Barriers may be studied and results used to design interesting photonic and optoelectronic devices. This article describes an energy-splitting phenomenon which is in accordance with experimental results that were already reported and involves 1-D photonic crystals [23] [24].

It has been already shown by our simulations that by modifying the photonic barriers' width the transmitted resonant states' total number can be controlled in these nanostructures. The resonant state describes here might be beneficial for detectors, optical filters, photonic-switching devices and other optoelectronic and photonic devices' new types' development. Now we are interested in Fibonacci series MQWs which could form the basis of future quantum computers. Model will be tested on existing structures and now testing on Fibonacci series multiple quantum well's structure and performance improved.

## ACKNOWLEDGEMENT

The University Grant Commission (UGC), India financially supported this work.

## REFERENCES

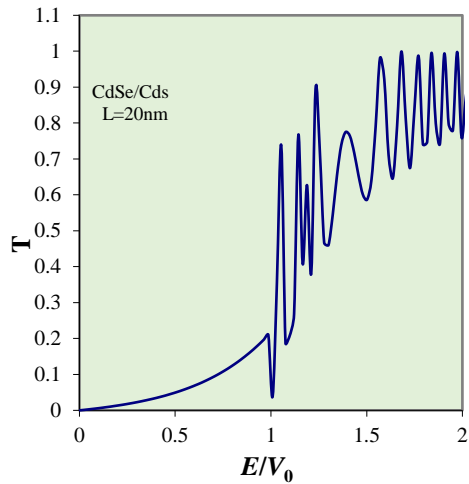

(a)

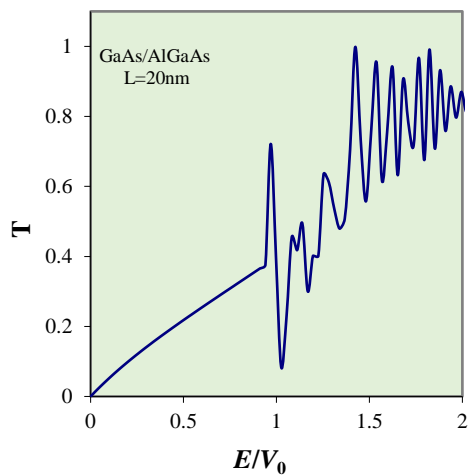

(b)

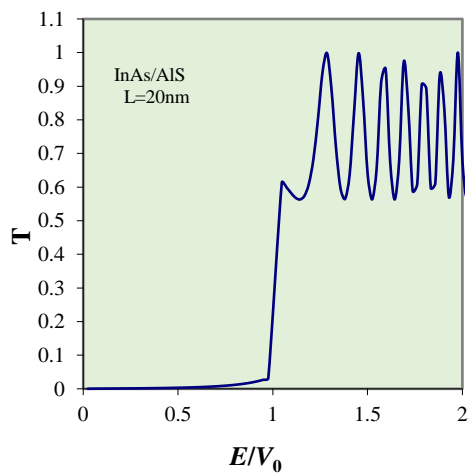

(c)

Fig.4. Variation of electrons' transmission coefficient ($T$) with normalized energy ($E/V_0$) for (a) CdS/CdSe (b) AlGaAs/GaAs (c) AlSb/InAs multiple quantum barriers (MQB) for barriers width ($L$) 20 nm

[1] Zhores I. Alferov, "Nobel Lecture: The Double Heterostructure Concept and Its Applications in Physics, Electronics, and Technology", *Review of Modern Physics*, Vol. 73, No. 2, pp. 697-703, 2002.

[2] Marco Zoli, "Instantonic Methods for Quantum Tunneling in Finite Size", *Brazilian Journal of Physics*, Vol.39, No. 4, pp. 1-12, 2009.

[3] Boris Mikhailovich Karnakov, Vladimir Pavlovich Krainov, "*WKB Approximation in Atomic Physics*", Springer, 2013.

[4] Chang Soo Park, Myung Geun Jeong, Sahng Kyoon Yoo and D.K. Park, "Double-Well Potential: The WKB Approximation with Phase Loss and Anharmonicity Effect", *Physical Review A*, Vol. 58, pp. 3443-3449, 1998.

[5] Vikas Maheshwari, Soma Shekhar Mali Patil, Neha Gupta and Rajib Kar, "Modified WKB Approximation for Fowler-Nordheim Tunneling Phenomenon in Nano- Structure based Semiconductors", *Proceedings of International Conference on Emerging Trends in Information Technology and Engineering*, pp. 446-457, 2020.

[6] Tao Pang, "A Numerical Method for Quantum Tunneling", *AIP Computers in Physics*, Vol. 9, pp. 602-612, 1995.

[7] R. Tsu and L Esaki, "Tunnelling in Finite Superlattice", *Applied Physics Letters*, Vol. 22, pp. 562-568, 1973

[8] D.K. Roy, "*Quantum Mechanical Tunnelling and Its Applications*", World Scientific Publishing, pp. 1-14, 1986.

[9] Hemendra Ghimire, P.V.V. Jayaweera, Divya Somvanshi, Yanfeng Lao and A.G. Unil Perera, "Review Recent

Progress on Extended Wavelength and Split- Off Band Heterostructure Infrared Detectors", *Micromachines*, Vol. 11, pp. 547-556, 2020.

[10] Deepika Tyagi, Huide Wang, Wenchuan Huang, "Recent Advances in Two Dimensional- Material-based sensing Technology Toward Health and Environmental Monitoring Applications", *Nanoscale*, Vol. 6, no. 6, pp. 1-12, 2020.

[11] Mitra Dutta and Michael A. Stroscio, "*Advanced Semiconductor Heterostructures*", World Scientific Publishing, 2003.

[12] Dongwoo Lee, David Blaauw, Dennis Sylvester, "Gate Oxide Leakage Current Analysis and Reduction for VLSI Circuits", *IEEE Transactions on Very Large Scale Integration (VLSI) Systems*, Vol. 12, No. 2, pp. 1-23, 2004.

[13] Juan C. Ranuarez, M.J. Deen and Chih Hung Chen, "A Review of Gate Tunneling Current in MOS Devices", *Microelectronics Reliability*, Vol. 46, No. 12, pp. 1939-1956, 2006.

[14] Victor Barsan, Mihaela Cristina Ciornei, "Semiconductor Quantum Wells with Ben Daniel-Duke Boundary Conditions: Approximate Analytical Results", *European Journal of Physics*, Vol. 38, No. 2, pp. 1-14, 2017.

[15] Sonia Bagumako, Ludovico Desplanque and Nicolas Wichmann, "100 nm AlSb/InAs HEMT for Ultra Low-Power Consumption, Low-Noise Applications", *Scientific World*, Vol. 2014, pp. 1-6, 2014.

[16] P.A. Alvi, "Transformation of Type-II inAs/AlSb Nanoscale Heterostructure into Type-I Structure and Improving Interband Optical Gain", *Physica Status Solidi*, Vol. 254, No. 5, pp. 1-24, 2017.

[17] Jian Feng and Min Xiao, "Lasing Action in Colloidal CdS/CdSe/CdS Quantum Wells", *Applied Physics Letters*, Vol. 87, No. 2, pp. 1-16, 2005.

[18] C. Bender and D. Hook, "Quantum Tunneling as a Classical Anomaly", *Journal of Physics A- Mathematical and Theoretical*, Vol. 44, No. 37, pp. 1-13, 2010.

[19] J. Vahedi and K. Nozari, "The Ramsauer-Townsend Effect in the Presence of the Minimal Length and Maximal Momentum", *Acta Physica Polonica A*, Vol. 122, pp. 1-16, 2012.

[20] C.S. Ho, J.J. Liou, Kuo-Yin Huang and Chin-Chang Cheng, "An Analytical Sub Threshold Current Model for Pocket-Implanted in MOSFETs", *IEEE Transactions on Electron Devices*, Vol. 50, No. 6, pp. 1475-1479, 2003.

[21] A. Bilhah and M.K. Das, "Influence of Doping on the performance of GaAs/AlGaAs QWIP for Long Wavelength Applications", *Opto Electronics Reviews*, Vol. 24, No. 1, pp. 1-14, 2016.

[22] Michael A. Demyanenko, Dmitry G. Esaev and A.I. Toropov, "AlGaAs/GaAs Quantum Well Infrared Photodetectors", Intechopen, Vol. 71266, pp. 1-13, 2018.

[23] Ivan S. Panyaev, Dmitry G. Sannikov, Nataliya N. Dadoenkova and Yuliya S. Dadoenkova, "Three-Periodic 1D Photonic Crystals for Designing the Photonic Optical Devices Operating in the Infrared Regime", *Applied Optics*, Vol. 60, pp. 1943-1952, 2021.

[24] W. Huaizhong, W. Zhanhua and B. Yang, "One-Dimensional Photonic Crystals: Fabrication, responsiveness and Emerging Applications in 3D Construction", *RSC Advances*, Vol. 6, pp. 4505-4520, 2016.